\documentclass[epj]{webofc}
\usepackage[varg]{txfonts}   
\begin{document}
%
%
\title{Results from the Pierre Auger Observatory}
%
%

\author{Esteban Roulet\inst{1}\fnsep\thanks{\email{roulet@cab.cnea.gov.ar} }, for the Pierre Auger Collaboration\inst{2}\fnsep\thanks{\email{ auger_spokespersons@fnal.gov }}
}

\institute{CONICET, Centro At\'omico Bariloche, Av. Bustillo 9500, Bariloche, Argentina
  \and
 Full author list and acknowledgments at: \url{http://www.auger.org/archive/authors_2018_09.html }
}

\abstract{%
  I describe some of the  results on ultrahigh-energy cosmic rays that have been obtained with the Pierre Auger Observatory. These include measurements of the spectrum, composition and anisotropies. Possible astrophysical scenarios that account for these results are discussed.
 }
\maketitle
\section{The Observatory}
The Pierre Auger Observatory was built to understand the nature and origin of the highest energy cosmic rays, i.e. those with energies above 1~${\rm EeV}\equiv 10^{18}$~eV. It is the largest existing observatory, covering an area of 3000~km$^2$ near the town of Malarg\"ue in Argentina, at a latitude of 35$^\circ$ South, and has been operating continuously since the year 2004. It consists of a surface array of water-Cherenkov detectors that can measure the lateral distribution of the air showers at ground level. This array is overlooked by 27 telescopes that detect, during clear moonless nights,  the fluorescence emitted by the air nitrogen molecules that get excited by the passage of the charged particles in the showers. These last observations allow us to determine the longitudinal profile of the electromagnetic component of the showers and in particular to measure the depth of its maximum development, known as $X_{\rm max}$. It also allow us to determine the energy of the showers through a calorimetric determination of the deposited energy, providing a clean calibration of the energy estimator obtained with the surface array. The Collaboration involves more than 400 scientists from 17 countries \cite{nim}. 

Some of the main results obtained up to now
have been, as  will be discussed here, the measurement of the high-energy suppression of the cosmic ray (CR) spectrum \cite{spectrum}, the observation of an increasingly heavier composition above the ankle \cite{compos} and the measurement of a  dipolar anisotropy at energies above 8~EeV which  indicates that  the CRs above this threshold have an extragalactic origin \cite{LS17}.
Many other results have also been obtained, such as the determination of the proton-air cross section at energies beyond those achieved at colliders \cite{xsect}, the studies of the muon content of the showers that provide clues about models of hadronic interactions \cite{muons} or the bounds on neutrino and gamma fluxes that exclude exotic scenarios for the cosmic ray origin \cite{nus,gammas}. Studies of large-scale anisotropies at different energies provided hints of a transition from an anisotropy of Galactic origin to one of extragalactic origin taking place at energies around the EeV \cite{LS11}. Also  hints of more localized anisotropies \cite{aniso15,sb}, on scales of 10$^\circ$--$20^\circ$, appear at the energies for which the spectrum starts to become strongly suppressed and  give the hope that with additional data, and with a better discrimination of the primary cosmic ray charges, the identification of the first individual ultrahigh energy CR sources may become feasible.  There is also a strong participation in the new multi-messenger era that exploits the combination of results from cosmic-ray observatories with those  from gravitational waves, high-energy neutrino and gamma-ray observatories \cite{gwnu}.

\section{The cosmic ray spectrum and composition}

\begin{figure}[t]
\centering
\includegraphics[scale=0.07]{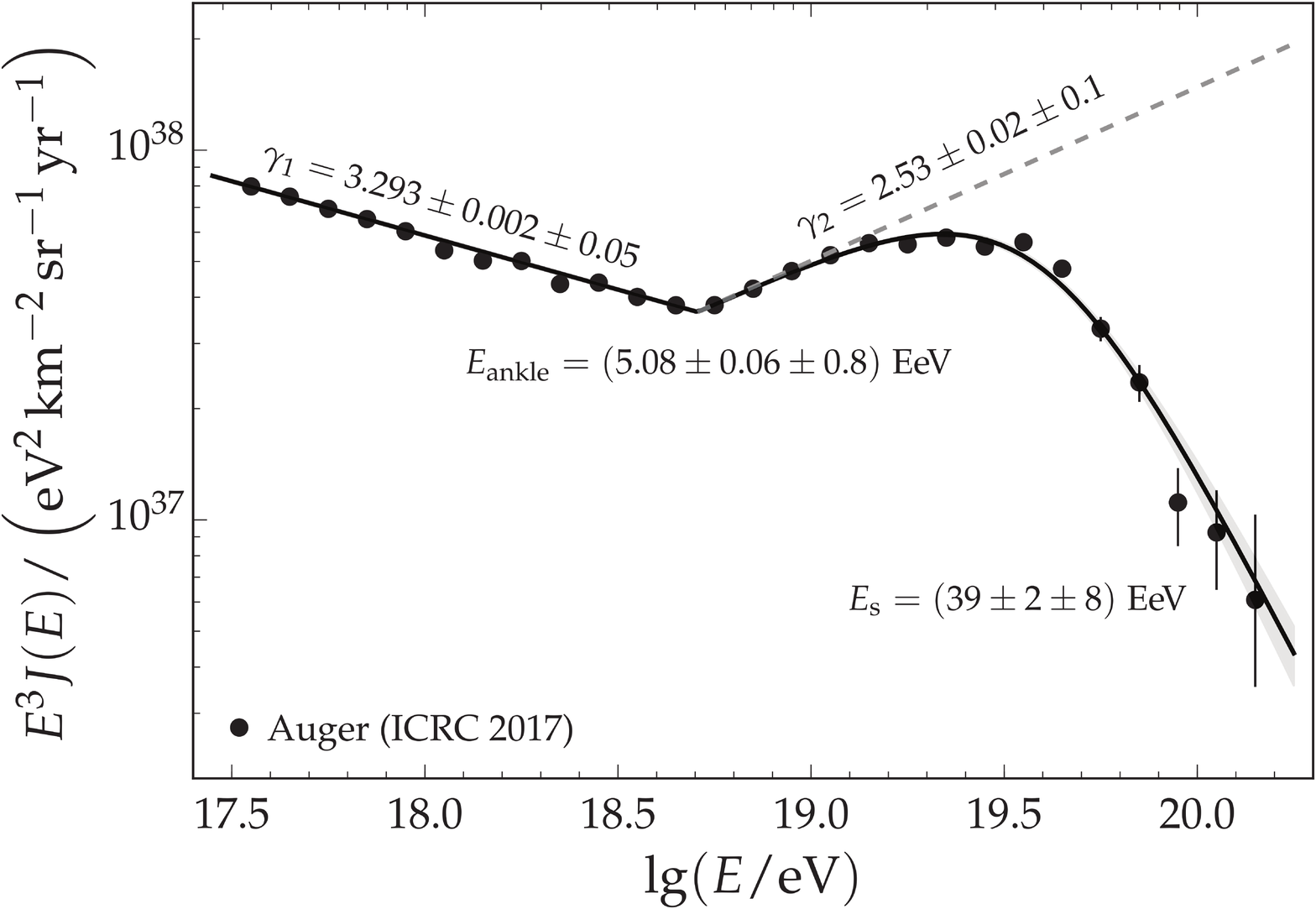}\hskip30pt\includegraphics[scale=0.31]{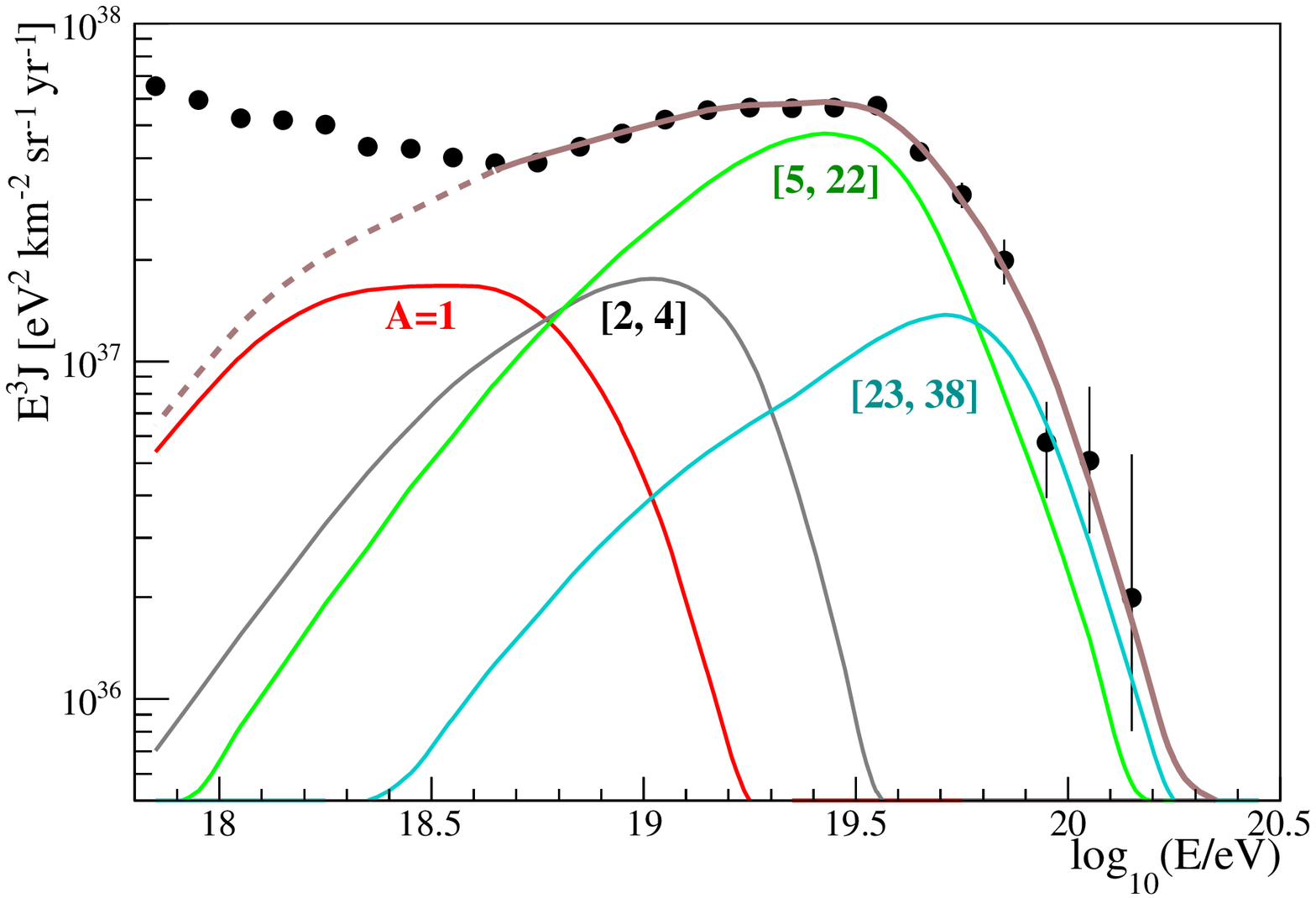}

\caption{Spectrum measured by the Auger Observatory and fitted parameters obtained using a broken power-law with a high-energy supression (see text) \cite{fenu}. The right panel shows the results obtained \cite{combined} by fitting the spectrum and $X_{\rm max}$ observations with a model of sources with a mixed composition of H, He, N and Si having a power-law spectra with a rigidity dependent cutoff. The individual curves depict the spectra at the Earth for different mass groups after the propagation and interactions with background radiation fields are taken into account. They correspond to the mass groups $A= 1$ (red), $2\leq A\leq 4$ (grey), $5\leq A \leq 22$ (green) and $23\leq A\leq 
38$ (cyan).
 }
\label{fig-1}      
\end{figure}

The left panel in figure~\ref{fig-1}  shows the measured CR flux, multiplied by $E^3$ in order to make it flatter, for energies above $3\times 10^{17}$~eV (the values below 1~EeV are obtained with a denser sub-array that covers an area of 23~km$^2$). The data points are fitted  to a broken power-law with a high-energy suppression, according to \cite{fenu}
\begin{eqnarray}
  J(E)&=& J_0\left(\frac{E}{E_{\rm ankle}}\right)^{-\gamma_1} \hskip140pt , E<E_{\rm ankle}\nonumber\\
  J(E)&=& J_0\left(\frac{E}{E_{\rm ankle}}\right)^{-\gamma_2}\left[1+\left(\frac{E_{\rm ankle}}{E_s}\right)^{\Delta\gamma} \right]
  \left[1+\left(\frac{E}{E_s}\right)^{\Delta\gamma} \right]^{-1} \hskip10pt, E\geq E_{\rm ankle}.
\end{eqnarray}
The so-called ankle break takes place at an energy $E_{\rm ankle}\simeq 5.1$~EeV, at which the spectrum hardens from a power law with index $\gamma_1\simeq 3.29$ to one with $\gamma_2\simeq 2.53$. At energies above $E_{\rm s}\simeq 39$~EeV the spectrum becomes strongly suppressed, being reasonably well fitted with a power-law index $\gamma_3\equiv \gamma_2+\Delta\gamma\simeq 5.0$. It is important to keep in mind that the overall uncertainty on the energy scale is 14\%.

\begin{figure}[ht]
\centering
\includegraphics[scale=0.70]{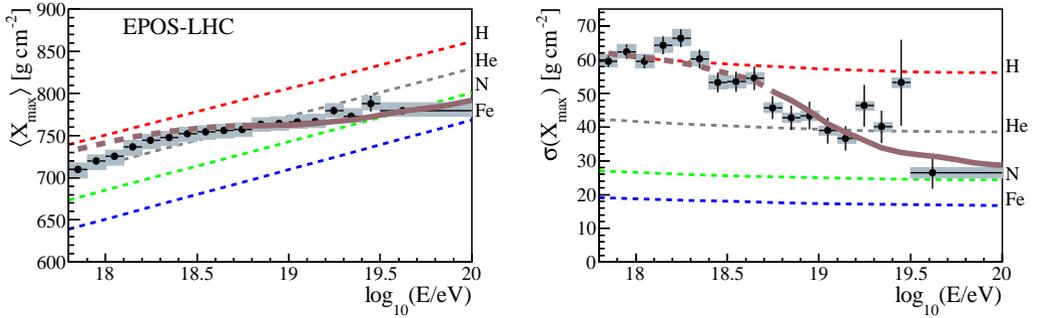}
\caption{Average and standard deviation of the $X_{\rm max}$ distribution, compared to those predicted assuming EPOS-LHC interactions for pure H (red  dashed lines), He (grey), N (green) and Fe (blue). Also shown with brown lines  are the predictions  \cite{combined} from the model with mixed composition illustrated in the right panel of figure~1, for which only the energy range where the brown lines are solid is included in the fit.
 }
\label{fig-2}     
\end{figure}

From the spectrum results alone it is however not possible to understand the causes responsible for the features observed, for instance if the ankle reflects the transition between a Galactic component at low energies towards a harder  extragalactic component at higher energies, or is instead reflecting the effects of a  feature in the attenuation of an extragalactic component (such as a threshold for pair production for CR protons interacting with the CMB), or something different. It is also not clear whether the suppression is the result of the GZK effect due to photo-pion production of primary protons, the photodisintegration of heavy nuclei like Fe, or just a cutoff in the maximum energies achievable at the sources. To gain insight into these issues it is important to measure the composition of the CRs, i.e. what is the distribution of the CR masses as a function of the energy. The best tool to achieve this is the study of the distribution of the $X_{\rm max}$ values, since showers of energy $E$ produced by nuclei of mass number $A$ can in first approximation be considered as a superposition of $A$ showers produced by nucleons of energy $E/A$. 
This means that for a given $E$ the lighter primaries produce showers that are on average more penetrating than those produced by heavier primaries. Moreover, the fluctuations on the depth of shower maximum, $\sigma(X_{\rm max})$, should decrease for increasing CR masses.

 The results obtained for the average and dispersion of the $X_{\rm max}$ distributions at different energies are displayed in figure~\ref{fig-2} (black dots with error bars) \cite{compos}. The main features are a break in the elongation rate (which is the change in $X_{\rm max}$ per decade of energy) which goes from about 80~g\,cm$^{-2}$ below an energy of about 2~EeV to a value of about 26~g\,cm$^{-2}$ above this energy. Since the elongation rate is expected to be about 56~g\,cm$^{-2}$ if the composition were to remain constant (the predictions for a single mass component are illustrated as dashed lines for the cases of pure H, He, N and Fe primaries), the observed values give  an indication that the average mass decreases as the energy approaches the EeV. One should recall that at 10$^{17}$~eV the CRs are believed to be predominantly of Galactic origin and are quite heavy. At a few EeV energies the $X_{\rm max}$ values suggest that the composition became light, with sizeable H and He components, while above these energies it should become increasingly heavier. The dispersion shown in the right panel is consistent with a light composition at EeV energies and becomes smaller  above 2~EeV, in agreement with a composition that becomes  heavier. Moreover, since a  mixture of two different mass components would lead to values of $X_{\rm max}$ with a larger dispersion  than any of the  components alone, the fact that $\sigma(X_{\rm max})$ becomes small above few EeV is an indication that there is no significant mixture  of CRs with disparate masses. 

 Figures~\ref{fig-1} (right panel) and \ref{fig-2} also display the results obtained by fitting the observations above 5~EeV with a scenario having a uniform distribution of identical sources with a mixture of four mass components (H, He, N and Si) \cite{combined}. Their spectra are assumed to be  power-laws of index $\gamma$  having a rigidity dependent  cutoff $R_{\rm cut}$, so that the spectra become exponentially suppressed above an energy  $E_{\rm cut}=Z_iR_{\rm cut}$, with $Z_i$ the corresponding atomic number of the nuclei of type $i$. The different nuclei are assumed to be produced with fractions $f_i$ (defined at energies smaller than $R_{\rm cut}$). Following the propagation of these nuclei from the sources up to the Earth and accounting for the interactions with the radiation backgrounds, the best fit to the spectra and $X_{\rm max}$ distributions lead to the results plotted in the figures, which correspond to values of $\gamma\simeq 1$ and $R_{\rm cut}\simeq 5$~EeV. The fitted composition fractions correspond to about 67\% He, 28\% N and 5\% Si, but note that the fit only includes data above 5~EeV and hence, given that the H cutoff obtained has a similar value, the H fraction is not really probed.

The value $\gamma\simeq 1$ obtained in the best fit  corresponds to a spectrum much harder than what is expected  in the Fermi diffusive shock acceleration scenarios, that predict values closer to $\gamma=2$. One of the main reasons for the small values of $\gamma$ inferred is the need to ensure that the heavier elements that dominate the spectrum at the highest energies become subdominant as the energy is decreased, so as to allow for a lighter composition and without leading to too large values of $\sigma(X_{\rm max})$. One possible way to obtain a hardening of the observed spectrum for rigidities below a certain value, even if the source spectrum is softer, is if the CRs diffuse in the turbulent extragalactic magnetic fields \cite{mr13}. If the density of sources is not too large so that the closest sources are at distances for which the CRs with rigidities below a certain value  would take  a time longer than the lifetime of the sources to reach the Earth, this `magnetic horizon' effect would lead to a strong suppression of the observed spectrum at low energies, and will hence effectively harden it. Indeed, a fit to the spectrum and composition results in scenarios that include intergalactic magnetic fields led to a best fit source power-law index $\gamma\simeq 1.6$ \cite{witko}, which is closer to the expectations.

\section{Large scale anisotropies}

Another crucial ingredient to understand the origin of the CRs, and to ultimately identify their sources, is the study of the distribution of their arrival directions to search for anisotropic signals. One of the main difficulties however is that, being charged, the CRs get deflected as they propagate through the Galactic and extragalactic magnetic fields and hence they do not point back to their sources when they are detected.  However, since the deflections get smaller for increasing energies, anisotropies may show-up for sufficiently high energies. One has to keep in mind that the deflection of an extragalactic proton due to the effects of the Galactic magnetic field alone is about 10$^\circ$ to 30$^\circ$ at 10~EeV, and in general it scales approximately as $Z/E$ for CRs with atomic number $Z$. The fact that above the ankle energy  the average CR charges appear to increase makes the  identification of individual sources more difficult, although anisotropies on large angular scales, such as a dipolar one, may still be observable. These kind of anisotropies are expected to result from the anisotropic distribution of the sources themselves, or also if some nearby sources contribute significantly and the CRs from them are sizeably deflected. Large angular scale anisotropies have already  been searched for a long time with the Auger observatory. A technique which is often adopted to obtain the equatorial dipolar component is the harmonic analysis in right ascension, exploiting  the almost uniform exposure achieved in this coordinate thanks to the Earth rotation. Different techniques exist to obtain the dipolar component along the Earth rotation axis, in particular one based on a harmonic analysis on the azimuth angle of the showers allows one to combine in a clean way different datasets, such as the so-called vertical one (including zenith angles $\theta\leq 60^\circ$) with the inclined one ($60^\circ < \theta\leq 80^\circ$, that includes events with $E>4$~EeV and relies on a different reconstruction than the one of the vertical sample). Many systematic effects have to be accounted for in order to obtain an accurate determination of the large scale anisotropies, such as  the small variations of the exposure with time (due e.g. to dead times), the atmospheric effects on the reconstructed energies (which can induce spurious daily and seasonal variations on the rates above a given threshold), the geomagnetic effects on the air showers (which  can induce a north-south asymmetry) or the slight tilt of the array towards the south-east.

\begin{figure}[t]
\centering
\includegraphics[scale=0.35]{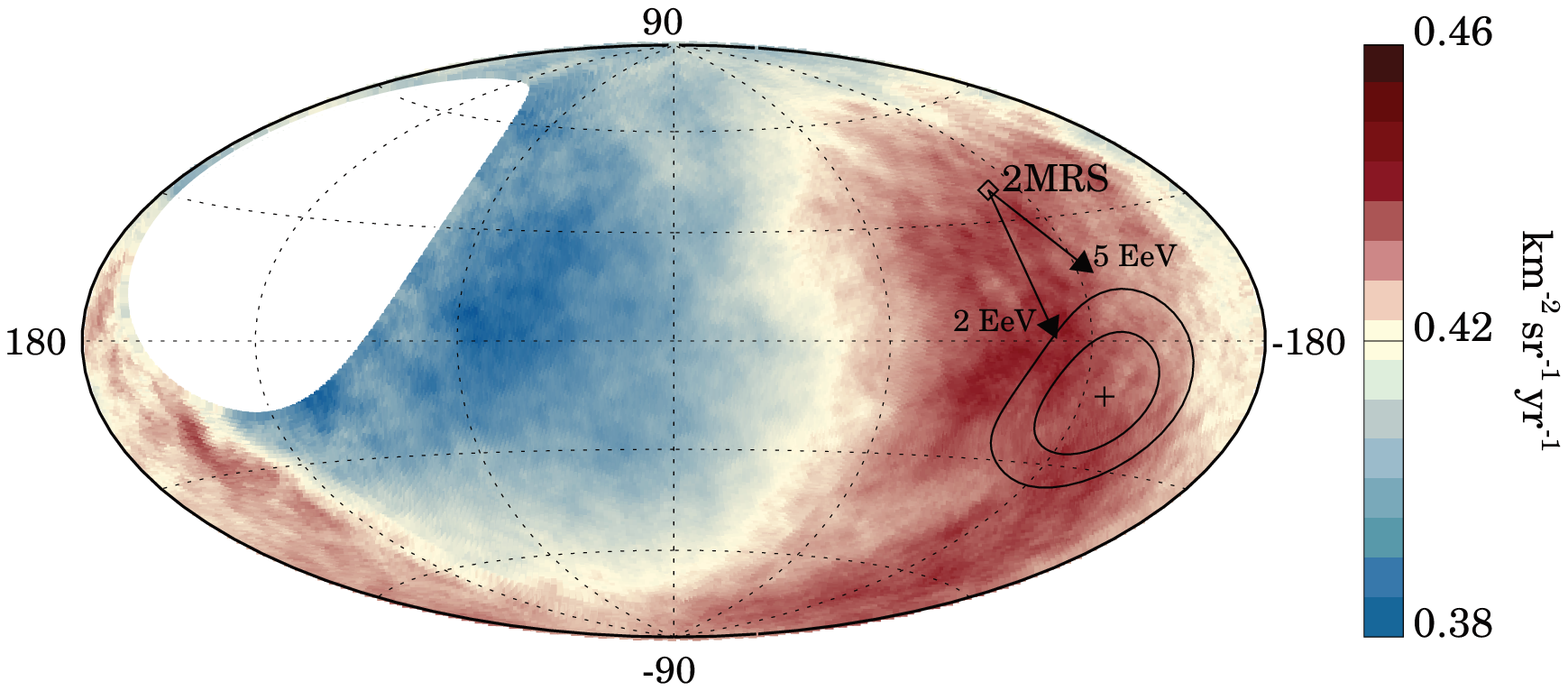}\hskip30pt\includegraphics[scale=0.28]{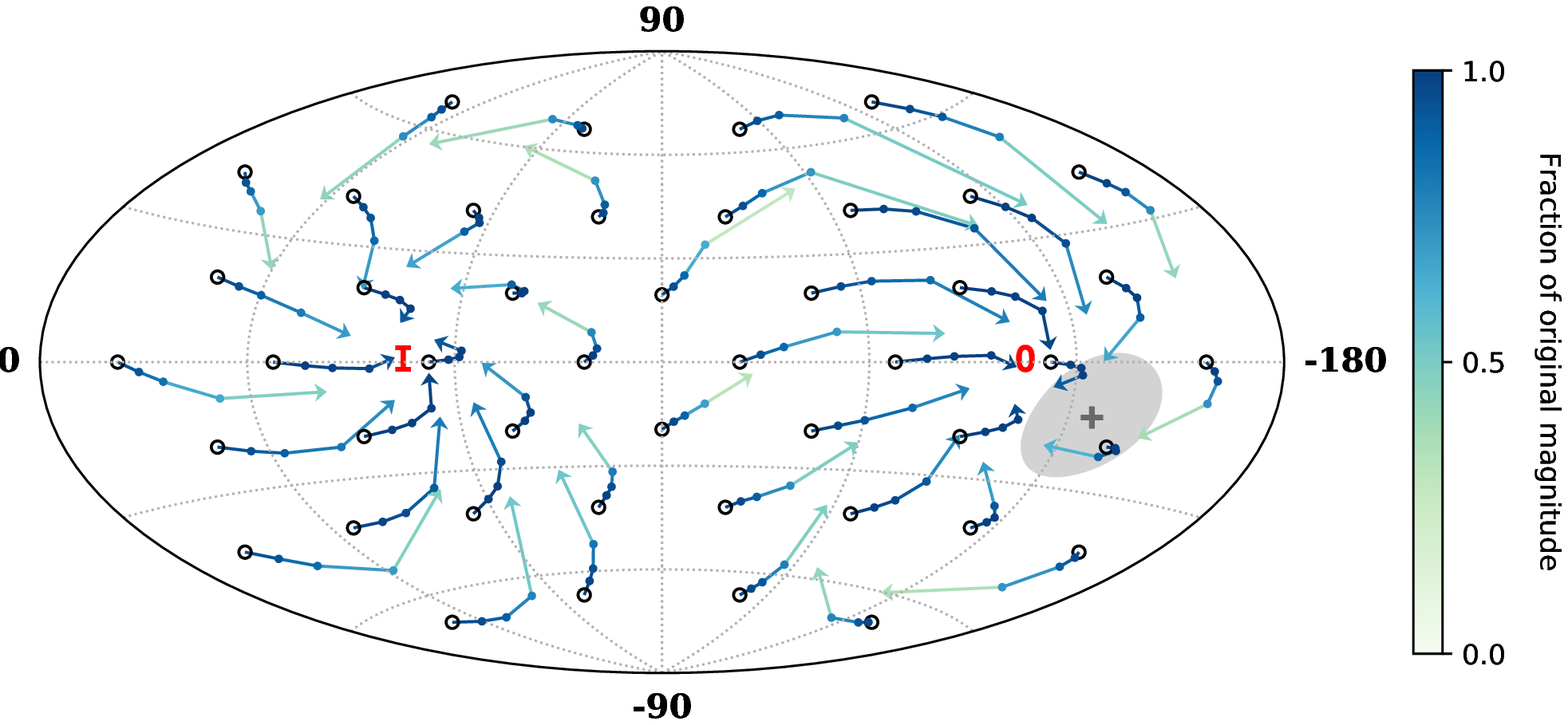}
\caption{Measured CR flux for $E>8$~EeV, averaged on top-hat windows of 45$^\circ$, in Galactic coordinates (left panel) \cite{LS17}. The direction of the reconstructed dipole direction is shown, together with the ellipses corresponding to the 68\% and 95\% CL regions.
   The right panel shows the changes in direction and amplitude of a dipolar extragalactic distribution after crossing the Galactic magnetic field, with circles being the original direction and the different points along the arrow corresponding to $E/Z= 32$~EeV, 16~EeV, 8~EeV and 4~EeV \cite{LS18}.}
\label{fig-3}      
\end{figure}

The amplitude of the equatorial dipole is constrained to be not much larger than the percent level at EeV energies, and this excludes a dominant Galactic origin for the light component of the CRs that is observed at these energies \cite{LS13}. Studying the energies above 4~EeV, for which the surface array is fully efficient up to zenith angles of $80^\circ$ (allowing to cover 85\% of the sky), and considering the two energy bins [4, 8]~EeV and $E\geq 8$~EeV \cite{LS15}, a first harmonic modulation in right ascension with a significance larger than 5$\sigma$ was found for  $E\geq 8$~EeV \cite{LS17}. The reconstructed CR flux in this energy range is shown in figure~\ref{fig-3} (left panel). It shows a clear dipolar pattern and the reconstructed 3D dipole has an amplitude of 6.5\%, pointing at Galactic coordinates 
$(\ell, b) = (233^\circ,\,-13^\circ)$. This direction  is  $\sim 125^\circ$ away from the Galactic center, suggesting that  this anisotropy has an extragalactic origin. Moreover, the dipole direction is not far from the direction of the outer spiral arm, which  is actually an attractor for the directions of an extragalactic dipole after the effects of the Galactic magnetic field deflections are taken into account \cite{LS18}. This is illustrated in the right plot in figure~\ref{fig-3}, where the  dipole directions that would be observed at the Earth are shown for different original dipole directions outside the Galaxy, considering the effects of the deflections due to the Galactic field, modeled as in \cite{jf12}. It is seen that the original directions that lie in the hemisphere of negative Galactic longitudes tend to be deflected towards the outer spiral arm (indicated with an {\bf O} in the map), while the other hemisphere is instead deflected towards the inner spiral arm (indicated as {\bf I}). The size of the deflections (and the resulting demagnifications of the dipole amplitude) depend on the CR rigidities. If the ultrahigh energy CR sources were to follow the distribution of the galaxies, one would expect that the anisotropy in the local distribution of galaxies should also reflect into an anisotropy in the CR arrival direction distribution. In particular, we show in the left panel of figure~\ref{fig-3} the direction towards the flux weighted dipole of the galaxies in the 2MRS catalog \cite{erdogdu}, which is dominated by the contribution from galaxies closer than about 100~Mpc. If the resulting CR dipole were to point in a similar direction, after taking into account the deflections of the Galactic magnetic field (illustrated with the arrows corresponding to two representative values of $E/Z$) the resulting direction would be not far from the one observed above 8~EeV.

\begin{figure}[t]
\centering
\includegraphics[scale=0.60,angle=0]{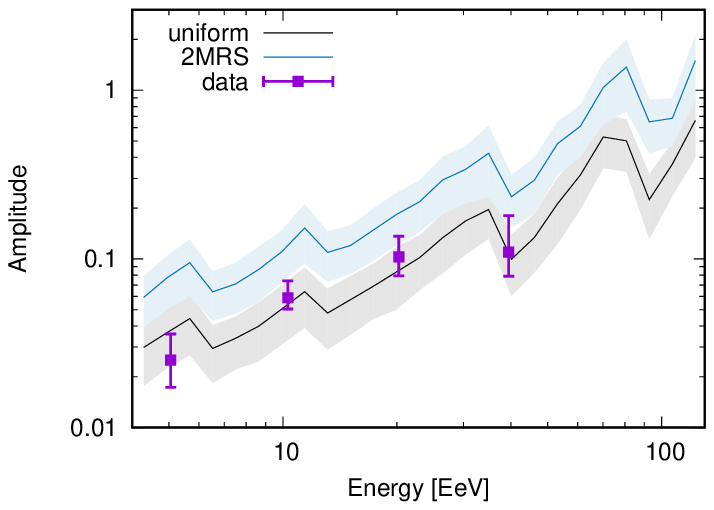}\hskip30pt\includegraphics[scale=0.40]{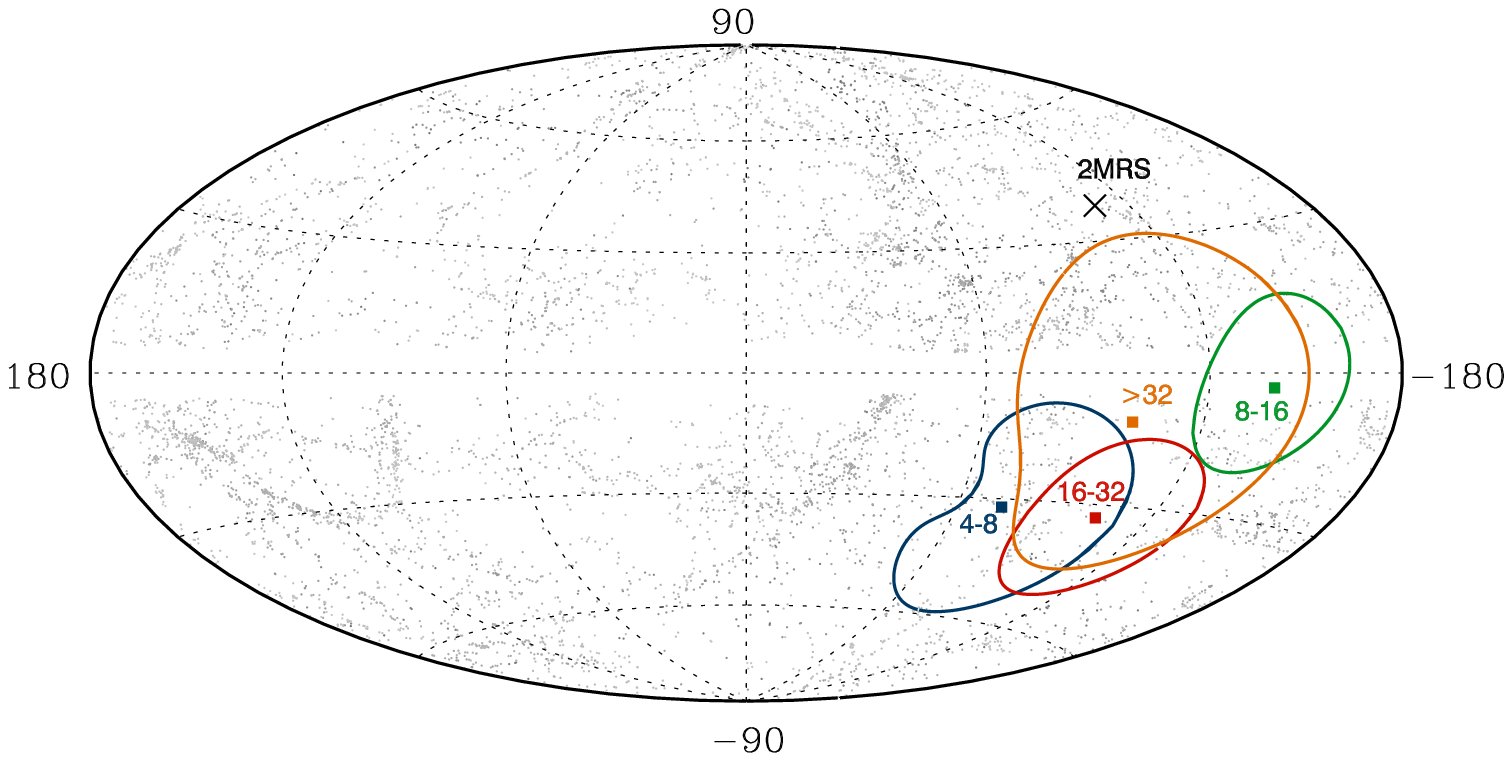}
\caption{Left panel: Energy dependence of the dipole amplitude, together with the expectations from different extragalactic source scenarios \cite{LS18}. Right panel: reconstructed dipole directions in the different energy bins.}
\label{fig-4}      
\end{figure}

In ref.~\cite{LS18} a detailed study of the energy dependence of this dipolar anisotropy has been performed, considering the energy bins [4, 8]~EeV, [8, 16]~EeV, [16, 32]~EeV and $E\geq 32$~EeV. The results are illustrated in figure~\ref{fig-4}. The left panel shows the 3D dipole amplitude as a function of the energy, as well as the expectations in scenarios in which extragalactic sources with a density of $10^{-4}$~Mpc$^{-3}$ are distributed uniformly or following the observed galaxy distribution \cite{hmr}. One can see that the overall trend of the amplitudes is quite consistent with the measured ones. The right panel shows the reconstructed dipole directions, which suggest an extragalactic origin for the anisotropies in all four bins. 

An extensive program of upgrades to the Pierre Auger Observatory is taking place at the moment, including the deployment of scintillators on top of each surface detector, small PMTs to increase the dynamic range,
radio antennas and new electronics. All this will help to measure more accurately the showers and should allow for mass discrimination of the primaries using the surface array, enhancing the capabilities to study the shower properties and to search for anisotropies restricting to the less deflected light component of the CR fluxes. Hence, the prospects for further improvements on our knowledge of the most energetic particles in nature are very promising.



\end{document}